\documentclass[a4paper,  12pt]{article}

\usepackage[T1]{fontenc}
\usepackage{amsmath}
\usepackage{amsthm}
\usepackage{amssymb}
\usepackage{graphicx}
\usepackage{float}
\usepackage{dsfont}
\usepackage{todonotes}
\usepackage{algorithm}
\usepackage{enumerate}
\usepackage{rotating}
\usepackage{tikz}
\usepackage{subcaption}
\usetikzlibrary{trees}
\usepackage{tabularx}

\newtheorem{theorem}{Theorem}
\newtheorem{lemma}[theorem]{Lemma}

\newtheorem{definition}[theorem]{Definition}
\newtheorem{proposition}{Proposition}

\newtheorem{remark}[theorem]{Remark}

\newcommand{\R}{\mathbb{R}}  
\newcommand{\E }{\mathbb{E}}  
\renewcommand{\P}{\mathbb{P}}  

\newcommand{\RSS}{{\operatorname{RSS}}}
\newcommand{\BIC}{{\operatorname{BIC}}}
\newcommand{\eqdis}{\stackrel{d}{=}}

\newcommand{\calA}{\mathcal{A}}
\newcommand{\calN}{\mathcal{N}}

\newcommand{\tr}{{\mathsf{T}}}

\title{Regularisation of regression trees by summation of $p$-values}
\author{Nils Engler\footnote{nils.engler@math.su.se, Department of Mathematics, Stockholm University, Sweden}, \, Mathias Lindholm\footnote{lindholm@math.su.se, Department of Mathematics, Stockholm University, Sweden}, \, Filip Lindskog\footnote{lindskog@math.su.se, Department of Mathematics, Stockholm University, Sweden} \, and Taariq Nazar\footnote{taariq.nazar@math.su.se, Department of Mathematics, Stockholm University, Sweden} }

\begin{document}

\maketitle

\begin{abstract}
The standard procedure to decide on the complexity of a CART regression tree is to use cross-validation with the aim of obtaining a predictor that generalises well to unseen data.  
The randomness in the selection of folds implies that the selected CART regression tree is not a deterministic function of the data. Moreover, the cross-validation procedure may become time consuming and result in inefficient use of training data. 
We propose a simple deterministic in-sample method that can be used for stopping the growing of a CART regression tree based on node-wise statistical tests. This testing procedure is derived using a connection to change point detection, where the null hypothesis corresponds to no signal.
The suggested $p$-value based procedure allows us to consider covariate vectors of arbitrary dimension and allows us to bound the $p$-value of an entire tree from above. Further, we show that the test detects a not too weak signal with a high probability, given a not too small sample size.

We illustrate our methodology and the asymptotic results on both simulated and real world data. Additionally, we illustrate how the $p$-value based method can be used to construct a deterministic piece-wise constant auto-calibrated predictor based on a given black-box predictor.
\end{abstract}



\section{Introduction}

When using binary-split regression trees in practice an important question is how to decide on the complexity of the constructed tree expressed in terms of, e.g., the number of binary splits in the tree, given data. Many applications focus on predictive modeling, where the objective is to construct a tree that generalises well to unseen data. The standard approach to decide on the tree complexity is then to use hold-out data and apply cross-validation techniques, see e.g.~\cite{hastie2009elements}. When constructing a tree by sequentially deciding on continuing to split,  adding new leaves to the tree in each step, cross-validation corresponds to a method for so-called ``early stopping''. When using a cross-validation-based early stopping rule, the constructed tree obviously depends on the hold-out-data for the different steps of the procedure. In particular, a randomised selection of hold-out data will inevitably result in the constructed tree being a random function of the data. This is not always desirable. Another aspect of using cross-validation is that this procedure may become time consuming and that data is not used in an efficient way, which is particularly important for smaller data sets. In the present paper a simple deterministic in-sample early stopping rule is introduced, which is based on $p$-values for whether to accept a binary split or not focusing specifically on CART regression trees, see e.g.~\cite{breiman1984classification}. For CARTs this type of reasoning has previously been considered with a focus on variable selection, see, e.g., \cite{shih2004variable}, where the focus is to use a $p$-value based criterion to counteract bias that is introduced if there are varying numbers of missing values across covariate dimensions. Related work not focusing on CARTs, that construct alternative binary split trees using a $p$-value based approach can be found in, e.g., \cite{hothorn2006unbiased}. The reason for focusing on CARTs is due to its widespread popularity amongst practitioners.

In order to explain the suggested tree-growing method, let $T_m$ denote a greedily grown optimal $L^2$ CART regression tree ($L^2$ refers to using a squared error loss function) with $m$ leaves (suppressing the dependence on covariates), see e.g.~\cite{breiman1984classification}. Input to the tree-growing method is a given sequence of nested regression trees $T_{m_1}, T_{m_2}, \ldots$, where $1=:m_{1} < m_{2} < \ldots$, i.e.~the first tree is simply a root node, each tree is a subtree of the next tree in the sequence, and no tree appears more than once.  
Note that $T_{m_j}$ and $T_{m_{j+1}}$ may differ by more than one leaf, i.e.~$m_{j + 1} - m_j \ge 1$. The tree-growing process starts from the root node $T_{m_1}$ by testing whether 
increasing the tree-complexity from $T_{m_1}$ to $T_{m_2}$ corresponds to a significant improvement in terms of the $L^2$ loss. 
If this is the case, the tree-growing process continues to test whether the tree-complexity should be increased from $T_{m_2}$ and $T_{m_3}$; otherwise the tree-growing process stops. 
If $m_{j+1} - m_j > 1$, all added splits are tested. 
The tree-growing process is
\begin{enumerate}[(i)]
	\item based on $p$-values so hypotheses and significance levels need to be specified, 

	\item an iterative procedure, possibly resulting in a large number of tests.
\end{enumerate}
Concerning (i): The null hypothesis, $H_0$, is that there is no signal in data. The alternative hypothesis, $H_A$, is that there is a sufficiently strong signal making a binary split appropriate. The significance level of the test can be seen as a subjectively chosen hyper-parameter, depending on the modeler's view on the Type I error. Concerning (ii): We cannot perfectly adjust for multiple testing, but it is possible to use Bonferroni arguments to bound the Type I error from above. By doing so the tree-growing process is stopped once the {\em sum} of the $p$-values is greater than the subjectively chosen overall significance level for testing the significance of the entire tree. If $m_{j+1} - m_j > 1$, then more than one $p$-value is added to the sum. Since the $p$-value based stopping rule relies on a Bonferroni bound, this tree-growing procedure will be conservative, tending to avoid fitting too large trees to the data. 

Relating to the previous paragraph it is important to recall that the tree-growing process is based on a given sequence of nested greedily-grown $L^2$ CART regression trees, and it is whether these binary splits provide significant loss improvements or not that is being tested. 
In order to compute a $p$-value for such a split it is crucial to account for that the split was found to be optimal in a step of the greedy recursive partitioning process that generated the tree.  
This is done by representing the tree-growing process as a certain change-point detection problem, building on results and constructions from \cite{yao1986asymptotic}. The usefulness of these results for change-point detection when analysing regression trees was noted in \cite{shih2004variable}. 

Using the change-point representation, the computed $p$-value for a node is used to investigate whether the optimal split amongst all possible splits results in a sufficiently strong loss reduction. In the current paper this is done one node at a time, properly taking all aspects of the optimality procedure into account under $H_0$. This is different from testing whether there is a change in mean between two mean predictors,  given that the covariate split has the strongest loss reduction of the splits considered. How this latter situation can be dealt with to account for so-called selective inference in a CART regression context is discussed in \cite{neufeld2022tree}.
Since the tree-growing process we consider is based on a given sequence of nested CART regression trees, we do not address variable selection issues. For more on CART regression trees and variable selection, see \cite{shih2004variable}. One can also note that the early stopping rule applied to CART suggested in the current paper is equivalent to pruning using the $p$-value rule from \cite{shih2004variable} if there are no missing values. The latter topic was not discussed or explored in \cite{shih2004variable}.

The $p$-values for loss improvements for a single locally optimally chosen binary split can be calculated exactly for small sample sizes $n$, but in practice large values for the sample size require approximations. In the current paper an asymptotic approximation is used, which is based on results from \cite{yao1986asymptotic} for a single covariate. A contribution of the current paper is to show that for covariate vectors of arbitrary dimension, the accuracy of the $p$-value approximation for a single binary split does not deteriorate substantially if we increase the dimension of the covariate vector. The $p$-value approximation for an entire tree, accounting for multiple testing issues, results in
\begin{enumerate}[(a)]
	\item a conservative stopping rule, given that the null hypothesis $H_0$ of no signal is true, i.e.~the tree-growing process will not be stopped too late, due to that we are using a Bonferroni upper bound, 
	\item that if the alternative hypothesis $H_A$ is true, the signal will be detected with a high probability if the sample size is sufficiently large. 
\end{enumerate}

Another aspect of the $p$-value based stopping criterion is that it can be thought of as a regularisation technique. In the present paper this fact is explored and the $p$-value based stopping criterion is related to common regularisation techniques such as AIC and covariance penalties. We quantify the degree of penalisation as a function of sample size.

So far we have focused on deterministic $p$-value based early stopping when constructing a single greedily grown optimal $L^2$ CART regression tree. In practice, however, trees are commonly used as so-called ``weak learners'' in boosting. The use of $p$-value based early stopping in tree based $L^2$ boosting is considered in Section~\ref{sec: numerical illustrations}. This is similar to the so-called ABT-machine introduced in \cite{huyghe2024boosting}, which uses another deterministic (not based on e.g.~cross-validation) stopping rule based on a sequence of nested trees obtained from so-called cost-complexity pruning, see \cite{breiman1984classification}. When applying $p$-value based stopping in a boosting setting, the inference interpretation becomes less clear, but it is still possible to view the procedure as a deterministic regularisation technique, which does not rely on time consuming cross-validation for training. This observation is not pursued further in the present paper.

Regression trees can be used as simple approximations of a given complex ``black-box'' predictor. The regression tree is an auto-calibrated predictor, see e.g.~\cite{kruger2021generic}, and it can typically be chosen without reducing the predictive accuracy much compared to the original black-box predictor. This is discussed in \cite{lindholm2023local}, where a CART regression tree is fitted to data using the black-box predictor as a single real-valued covariate. The fact that the use of cross-validation implies a random (CART regression tree) predictor was criticised in \cite{wuthrich2024isotonic}, where isotonic regression was suggested as a preferred approach due to its non-randomness. 
The selection of the CART regression tree using the $p$-value method studied in the current paper does not use cross-validation and is hence an alternative to isotonic regression, and does not require the monotonicity assumptions of isotonic regression. 
In Section~\ref{sec: numerical illustrations}, we illustrate, based on real data and a given tree-based gradient boosting machine (GBM) predictor, see \cite{friedman2001greedy}, how our $p$-value based method gives a simple and accurate approximation of a complex ``black-box'' predictor.

The above discussion has focused on how to bound the Type I-error for a full tree, starting from a given CART regression tree. An alternative tree building procedure is to instead apply the $p$-value based stopping rule recursively, node by node, without controlling the tree's overall Type I-error. This is the type of procedure used in \cite{hothorn2006unbiased}. This approach will not be discussed in detail in the current paper, but it can be noted that this approach will tend to be more liberal (result in a larger tree) than the approach suggested by us that applies a Bonferroni approximation for controlling the overall Type I-error.

The predictive performance of a model is typically assessed based on hold out data, and this is done in the numerical illustrations in the present paper. There is no conceptual conflict in doing this for the $p$-value based method, since the meaning of the subjectively chosen nominal confidence level will remain the same when applied to the full training data. This differs from common approaches for selection of hyper-parameters such as parameters needed for pruning a regression tree.

Although we focus only on CART regression trees, one may, of course, consider other types of regression trees and inference based procedures to construct trees. For more on this, see e.g.\ \cite{hothorn2006unbiased}. 

\smallskip

{\bf Our main contribution.}
Given an arbitrary sequence of nested $L^2$ CART regression trees, grown by greedy optimal recursive partitioning, we provide an easy-to-use deterministic stopping rule for deciding on the regression tree with suitable complexity. We allow for covariate vectors of arbitrary dimension and the stopping rule is formulated in terms of an easily computable upper bound for the $p$-value corresponding to testing the hypothesis of no signal. Because of the upper bound, the stopping rule is conservative. However, we provide a theoretical guarantee that if there exists signal, then we will detect the existence of this signal if the sample size is sufficiently large. In particular, it is unlikely that we will stop the tree-growing process too early. The asymptotic theoretical guarantee is confirmed by numerical experiments.  All code is available at GitHub.\footnote{\label{fnote: github}Source code avaiable at \texttt{https://github.com/taariqnazar/pval-regularized-trees}}

\smallskip

{\bf Organisation of the paper.}  
The remainder of the paper is structured as follows.
Section \ref{sec: regression trees} introduces $L^2$ CART regression trees and sequences of nested such trees. 
Section \ref{sec:stopping_rule} presents and motivates the suggested stopping rule. 
Section \ref{sec:change_point_detection} describes that the stopping rule naturally leads to considering a change-point-detection problem and presents theoretical results that guarantee statistical soundness of our approach for large sample size.  
Section \ref{sec:regularisation_techniques} compares, for a single split, our approach to well-established regularisation techniques.  
Section \ref{sec: numerical illustrations} provides a range of numerical illustrations, both in order to clarify the finite-sample performance of our approach and also to illustrate useful applications for selection of a single CART regression tree without the use of cross-validation. 
The proofs of the main results are found in the appendix.

\section{Regression trees}\label{sec: regression trees}

The Classification and Regression Tree (CART) method was introduced in the 1980s and uses a greedy approach to build a piecewise constant predictor based on binary splits of the covariate space, one covariate at a time, see e.g.\ \cite{breiman1984classification}. If we let $x$ be a $d$-dimensional covariate vector with $x \in \mathbb{X} \subseteq \mathbb{R}^d$, a regression tree with $m$ leaves can be expressed as
\begin{align}\label{eq: CART simple}
    x\mapsto T_m(x) := \sum_{k=1}^{m} \zeta_k \mathds{1}_{\{x \in \mathbb{A}_k \}},
\end{align}
where $\zeta_k \in \mathbb{R}$, where $\mathbb{A}_k \subset \mathbb{X}, \cup_{k = 1}^m \mathbb{A}_k = \mathbb{X}$, and where $\mathds{1}_{\{x \in \mathbb{A}_k \}}$ is the indicator such that $\mathds{1}_{\{x \in \mathbb{A}_k \}}=1$ if $x \in \mathbb{A}_k$, and $0$ otherwise. For binary split regression trees, having $m$ leaves corresponds to having made $m-1$ binary splits. 

The construction of a CART regression tree is based on recursive greedy binary splitting. A split is decided by, for each covariate dimension $j$, considering the best threshold value $\xi$ for the given covariate dimension, and finally choosing to split based on the best covariate dimension and the associated best threshold value. Splitting the covariate space $\mathbb{X}$ based on the $j$th covariate dimension and threshold value $\xi$ corresponds to the two regions 
\begin{align*}
\mathbb{R}_{\text{left}}(j, \xi) = \{x \in \mathbb{X} : x_j \le \xi \}, \quad \mathbb{R}_{\text{right}}(j, \xi) = \{x \in \mathbb{X} : x_j > \xi\}. 
\end{align*}
The CART algorithm estimates a regression tree by recursively minimising the empirical risk based on the observed data $(Y^{(1)}, X^{(1)}), \ldots, (Y^{(n)}, X^{(n)})$ that are independent copies of $(Y, X)$, where $Y$ is a real-valued response variable and $X$ is a $\mathbb{X}$-valued covariate vector.  
When using the $L^2$ loss and considering a split w.r.t.~covariate $j$, this means that we want to minimise
\begin{align}\label{eq: CART minimisation}
    \sum_{i : X^{(i)} \in \mathbb{R}_{\text{left}}(j,\xi)} (Y^{(i)} - \overline{Y}_{\text{left}}(j,\xi))^2 + \sum_{i  : X^{(i)} \in \mathbb{R}_{\text{right}}(j,\xi)} (Y^{(i)} -  \overline{Y}_{\text{right}}(j,\xi))^2,
\end{align}
where $\overline{Y}_{\text{left}}(j,\xi)$ is the average of all $Y^{(i)}$ for which $X^{(i)} \in \mathbb{R}_{\text{left}}(j,\xi)$, and similarly for $\overline{Y}_{\text{right}}(j,\xi)$. 
A regression tree with a single binary split w.r.t.~covariate $j$ and threshold value $\xi$ is therefore 
\begin{align*}
    T_2(x) = \overline{Y}_{\text{left}}(j,\xi) \mathds{1}_{\{x \in \mathbb{R}_{\text{left}}(j,\xi)\}} + \overline{Y}_{\text{right}}(j,\xi) \mathds{1}_{\{x \in \mathbb{R}_{\text{right}}(j,\xi)\}}.
\end{align*}

In order to ease notation, it is convenient to fix a covariate dimension index $j$ and considered the the ordered pairs $(Y^{(1)},X^{(1)}), \dots, (Y^{(n)},X^{(n)})$ of $(Y,X)$, where we assume ordered covariate values $X^{(1)}_j\leq \dots \leq X^{(n)}_j$ and that the response variables appear in the order corresponding to the size of the covariate values. Hence, $(Y^{(1)},X^{(1)})$ satisfies $X^{(1)}_j=\min_i X^{(i)}_j$, etc. A different choice of index $j$ would therefore imply a particular permutation of the $n$ response-covariate pairs. By suppressing the dependence on $j$, this allows us to introduce
\begin{align}\label{eq:S_sums}
S_{\leq r} :=\sum_{i=1}^{r}(Y^{(i)}-\overline{Y}_{\leq r})^2, \quad  S_{> r}:=\sum_{i=r+1}^{n}(Y^{(i)}-\overline{Y}_{> r})^2,  \quad S:=S_{\leq n}, 
\end{align} 
where 
\begin{align*}
\overline{Y}_{\leq r}:=\frac{1}{r}\sum_{i=1}^{r}Y^{(i)}, \quad \overline{Y}_{> r}:=\frac{1}{n-r}\sum_{i=r+1}^{n}Y^{(i)}.
\end{align*}
That is, minimisation of \eqref{eq: CART minimisation} is equivalent to minimising $S_{\leq r}+S_{> r}$
with respect to $r$, or alternatively we can consider maximising the relative $L^2$ loss improvement, given by
\begin{align}\label{eq: CART split improvement}
	\frac{S - (S_{\leq r} +  S_{> r})}{S}.
\end{align}
Further, note that unless we build balanced trees with a pre-specified number of splits we need to add a stopping criterion  to the tree-growing process. The perhaps most natural choice is to consider a threshold value, $\vartheta$, say, such that the recursive splitting only continues if the optimal $r$, denoted $r^*$, for the optimally chosen covariate dimension $j^* \in \{1, \ldots, d\}$ satisfies
\begin{align}\label{eq: CART threshold}
	\frac{S - (S_{\leq r^*} +  S_{> r^*})}{S} > \vartheta.
\end{align}
This means that the threshold parameter $\vartheta$ functions as a hyper-parameter. In particular, if we let $T_m$ denote a recursively grown $L^2$ optimal CART regression tree with $m$ leaves created using the threshold parameter $\vartheta$, then for any subtree $T_m$ of $T_{m'}$, $m < m'$, the corresponding threshold parameters satisfy $\vartheta > \vartheta'$. Threshold parameters $\vartheta_1 > \vartheta_{2} > \ldots > \vartheta_{\tau}$ generate a sequence of nested trees $T_{m_1}, T_{m_2}, \ldots, T_{m_\tau}$ with $m_1\leq m_2 \leq \ldots \leq m_{\tau}$. In applications we will consider sequences $\vartheta_1 > \vartheta_2 > \ldots$ such that $1=m_1<m_2<\ldots$. 
Note that such a decreasing sequence of threshold parameters will not necessarily result in a sequence of nested trees that only increases by one split at a time. 

One procedure to construct a sequence of nested trees is to first pick $\vartheta = 0$ and build a maximal CART regression tree, which is pruned from the leaves to the root. One such procedure is the cost-complexity pruning introduced in \cite{breiman1984classification}, which likely will lead to a sequence of nested trees where more than one leaf is added in each iteration. For more on this, see Section~\ref{sec: cost-complexity pruning}.

The threshold parameter $\vartheta$ controls the complexity of the tree that is constructed using recursive binary splitting, but it is not clear how to choose $\vartheta$. One option is to base the choice of $\vartheta$ on out-of-sample validation techniques, such as cross-validation. The drawback with this is that the tree construction then becomes random: given a fixed dataset repeated application of the procedure may generate  different regression trees. We do not want a procedure for constructing regression trees to have this feature. The focus of the current paper is to start from a sequence of nested greedy binary split regression trees, from shallow to deep, and use a particular stopping criterion to decide when to stop the greedy binary splitting in the tree-growing process. The stopping criterion is based entirely on the data used for building the regression trees and is a deterministic mapping from the data to the elements in the sequence of regression trees. 

\subsection{The stopping rule}\label{sec:stopping_rule}

Our approach relies on that all binary splits in the sequence of nested regression trees have been chosen in a greedy optimal manner. That is, if we consider an arbitrary binary split in the sequence of nested trees, the reduction in squared error loss is given by the statistic
\begin{align}\label{eq: loss improvement total}
	U_{\max} := \max_{1\leq j\leq d}U_j, \quad U_j:=\max_{1\leq r \leq n-1}\frac{S-(S_{\leq r} + S_{>r})}{S}, 
\end{align}
where the sums $S_{\leq r}$ and $S_{>r}$ depend on $j$ because of the implicit ordering of the terms as outlined above, see \eqref{eq:S_sums}. 
Given any sample size $n$ and any observed value $u_{\text{obs}}$ for the test statistic $U_{\max}$ we easily compute, under the null hypothesis of no signal, an upper bound $p_{\text{obs}}\geq \P_{\calN}(U_{\max}>u_{\text{obs}})$, where the subscript $\calN$ emphasizes the null hypothesis. Therefore, for a regression tree $T_m$ resulting from $m-1$ binary splits, it holds that 
\begin{align*}
\P_{\calN}\bigg(\bigcup\limits_{k=1}^{m-1} \big\{U_{\max,k}>u_{\text{obs},k}\big\}\bigg)
\leq \sum_{k=1}^{m-1}\P_{\calN}(U_{\max,k}>u_{\text{obs},k})
\leq \sum_{k=1}^{m-1}p_{\text{obs},k}. 
\end{align*}  
Note that the summation is over all $m-1$ splits (or internal nodes) of the tree with $m$ leaves. We emphasize that, for every binary split $k$, $u_{\text{obs},k}$ is observed and $p_{\text{obs},k}$ is easily computed from $u_{\text{obs},k}$.    
If for a pre-chosen tolerance $\delta\in (0,1)$ close to zero, 
\begin{align}\label{eq:continuation}
\sum_{k=1}^{m-1}p_{\text{obs},k}\leq \delta,
\end{align}
then we conclude that the event $\cup_{k=1}^{m-1}\{U_{\max,k}>u_{\text{obs},k}\}$ is very unlikely and we reject the null hypothesis of no signal. Consequently, we proceed by considering the next, larger, regression tree $T_{m'}$, $m'>m$, in the sequence of nested regression trees. If, when considering the regression tree $T_{m'}$ we find that 
\begin{align}\label{eq:stopping_criterion}
\sum_{k=1}^{m'-1}p_{\text{obs},k}>\delta,
\end{align} 
then the procedure stops and the previous regression tree $T_m$ is selected as the optimal regression tree. The confidence level $\delta$ can be thought of as a subjectively chosen hyper-parameter, chosen to reflect the modeler's view on an acceptable level of Type I-error. 

Since we consider an upper bound for the probability (under the null hypothesis) of the event $\cup_{k=1}^{m-1}\{U_{\max,k}>u_{\text{obs},k}\}$ and since we consider upper bounds $p_{\text{obs},k}$ for the probabilities of the events $U_{\max,k}>u_{\text{obs},k}$, we are more likely to stop -- observe that \eqref{eq:stopping_criterion} holds -- compared to a hypothetical situation where the probability of the event $\cup_{k=1}^{m-1}\{U_{\max,k}>u_{\text{obs},k}\}$ could be computed and were found to exceed the tolerance level $\delta$. Hence, our stopping criterion is conservative. We therefore have to be concerned with the possibility of a too conservative stopping criterion. However, it is shown in Proposition \ref{thm:pvaluetozero} below that under an alternative hypothesis of a sufficiently strong signal, the computable upper bound $p_{\text{obs}}$ for the true $p$-value is very small. Hence, our stopping criterion is not too conservative.  
 
\subsection{Change point detection for a single binary split}\label{sec:change_point_detection}

The question of whether a candidate binary split should be rejected or not can be phrased as a change-point-detection problem.  
This observation has been made already in \cite{shih2004variable}, where the aim was to target inference based variable selection. The idea here is to make inference on squared error loss reduction, where a significant loss reduction translates into not rejecting a split, hence continuing the tree-growing process. This approach builds on the analysis of change-point detection from \cite{yao1986asymptotic} that uses a scaled version of \eqref{eq: loss improvement total} according to 
\begin{align*}
U^{(n)}_j:=\max_{1\leq r \leq n-1}\frac{S - (S_{\leq r} + S_{>r})}{S/n}, \quad U^{(n)}_{\max}:=\max_{1\leq j\leq d}U^{(n)}_j, 
\end{align*}
where the dependence of $U^{(n)}_j$ on $j$ is implicit in the order of $Y^{(1)},\dots,Y^{(n)}$ which determines the sums of squares $S_{\leq r}$ and $S_{>r}$, as before. That is, the optimal candidate change point w.r.t.~covariate dimension $j$ is expressed in terms of the statistic $U_j^{(n)}$, which, hence, is identical to a candidate split point.

The test for rejecting a candidate split is based on the null hypothesis saying that observing $X$ gives no information about $Y$. 
The null hypothesis corresponds to a simple model $\calN$ for $(Y,X)$.  
\begin{definition}[Null hypothesis, $H_0$]
For model $\calN$, $Y$ and $X$ are independent and $Y$ is normally distributed: 
there exist $\mu\in\R$ and $\sigma^2\in (0,\infty)$ such that  
\begin{align*}
\P_{\calN}(Y\in \cdot \mid X)=\P_{\calN}(Y\in \cdot)=N(\mu,\sigma^2).
\end{align*} 
\end{definition}
When considering a nested sequence of binary regression trees, $U^{(n)}_{\max}$ is the random variable whose outcome is the observed test statistic for a single candidate binary split. 
Under the null hypothesis, the common distribution of the statistics $U^{(n)}_{1},\dots,U^{(n)}_{d}$ does not depend on $\mu$ and $\sigma$. 
Hence, under the null hypothesis, the distribution of $U^{(n)}_{\max}$ does not depend on $\mu$ and $\sigma$. 
Clearly, 
\begin{align}\label{eq: p-value approximation U}
	\P_{\calN}\big(U^{(n)}_{\max}>u\big)&=\P_{\calN}\Big(\cup_{j=1}^{d}\{U^{(n)}_j>u\}\Big)\leq d\P_{\calN}\big(U^{(n)}_j>u\big)
\end{align}
which does not depend on $j$ since the probability is evaluated under the null hypothesis. 
We approximate the tail probability $\P_{\calN}(U^{(n)}_j>u)$ by $p_n(u)$, where 
\begin{align}\label{eq: p-value approximation U_j}
	p_{n}(u):=1-\Phi\bigg(u^{1/2}-\frac{\ln_3(n)+\ln(2)}{(2\ln_2(n))^{1/2}}\bigg)^{2\ln(n/2)},
\end{align}
where $\ln_k(n)$ corresponds to the $k$ times iterated logarithm, e.g.~$\ln_2(n)=\ln(\ln(n))$.  
The approximation $p_n(u)$ from \eqref{eq: p-value approximation U_j} corresponds to Eq.~(2.5) on p.~345 in \cite{yao1986asymptotic}. 
The true $p$-value is the function $u\mapsto \P_{\calN}\big(U^{(n)}_{\max}>u\big)$ evaluated at the observed value for $U^{(n)}_{\max}$. 
The true $p$-value is approximated from above by 
\begin{align}\label{eq: p-value bound}
	P_{\max}^{(n)}:=dp_{n}(U^{(n)}_{\max}). 
\end{align}
We emphasise that given an observation $u_{\text{obs},k}$ of $U^{(n)}_{\max}$, $p_{\text{obs},k}$ is the observed outcome of $P_{\max}^{(n)}$. We emphasise that $H_0$ is a statement about the true data generating process. In practice there will likely be dependence between the response variables corresponding to a specific region in the tree. This means that rejecting $H_0$ may be due to different reasons, both related to dependent responses and due to mean functions different from that specified by $H_0$.

If the true signal is not too weak, which means that the conditional expectation of $Y$ given $X$ should fluctuate sufficiently in size,  
then for any significance level we want to reject the null hypothesis in a setting with sufficiently large sample size $n$. 
In order to make the meaning of this statement precise, and in order to verify it, we must consider the alternative hypothesis as a sequence of hypotheses indexed by the sample size $n$. 
The alternative hypothesis corresponds to a sequence of models $\calA=(\calA^{(n)})$.  

\begin{definition}[Alternative hypothesis, $H_A$]\label{def:alt_hyp}
For the sequence of models $(\calA^{(n)})$ there exist $j\in\{1,\dots,d\}$, $\xi\in\R$, $t_0\in (0,1)$, $\sigma^2\in (0,\infty)$ and  
$\mu_l,\mu_r\in\R$, $\mu_l\neq\mu_r$, such that for all $n$ 
\begin{align*}
&\P_{\calA^{(n)}}(X_j\leq \xi)=t_0, \\
&\P_{\calA^{(n)}}(Y\in \cdot \mid X_j=x)=N(\mu_l\mathds{1}_{\{x<\xi\}}+\mu_r\mathds{1}_{\{x\geq \xi\}},\sigma^2),  
\end{align*}
where $ |\mu_r-\mu_l|=\sigma\theta_n>0$. 
The sequence $\theta_n$ satisfies 
\begin{align}\label{eq:stepsize}
\theta_n=\frac{(2\ln_2(n))^{1/2}+\eta_n}{n^{1/2}(t_0(1-t_0))^{1/2}}
\end{align}
for some increasing sequence $\eta_n$ with $\lim_{n\to\infty}\eta_n=\infty$ and $\limsup_{n\to\infty}\theta_n<\infty$.   
\end{definition}
The requirement under the alternative hypothesis of a shift in mean of size $\sigma\theta_n$ says that the amplitude of the signal is allowed to decrease towards zero with $n$, but not too fast. We could consider $\theta_n=n^{-r}$ for some $r<1/2$. We may also consider a constant signal amplitude $\theta$. However, that situation is not very interesting since such a signal should eventually be easily detectable as the sample size $n$ becomes very large. 
The expression for $\theta_n$ in \eqref{eq:stepsize} comes from \cite{yao1986asymptotic} (Eq.~(3.2) on p.~347) and corresponds to an at least slightly stronger signal compared to what was considered in \cite{yao1986asymptotic} ($\eta_n\to\infty$ instead of $\eta_n=\eta+o(1)$). 

We want to show that under the alternative hypothesis we will reject the null hypothesis with a probability tending to one. The null hypothesis is not rejected at significance level $\varepsilon>0$ if $P_{\max}^{(n)}>\varepsilon$. We want to show that under the alternative hypothesis, the probability of falsely not rejecting the null hypothesis is very small. More precisely, we show the following:
\begin{proposition}\label{thm:pvaluetozero} 
$\lim_{n\to\infty}\P_{\calA^{(n)}}(P_{\max}^{(n)}>\varepsilon)=0$ for every $\varepsilon>0$.   
\end{proposition}
\noindent The proof of Proposition~\ref{thm:pvaluetozero} is given in the Appendix.

Similar to the discussion following \eqref{eq: p-value bound}, the alternative hypothesis $H_A$ is a statement about the true underlying data generating process. In practice, when creating a tree recursively, dependence between response variables within node regions may be observed. This dependence is in general hard to characterise and exploit, and Proposition~\ref{thm:pvaluetozero} should be seen as a stylised result indicating soundness of the proposed method.

\section{Relation to classical regularisation techniques}\label{sec:regularisation_techniques}

The focus of this section is on a single binary split. Let $T_2(X)$ denote an optimal binary-split CART regression tree with a single split, and let $T_1(X)$ denote the root tree of $T_2(X)$. Based on the notation in Section~\ref{sec: regression trees} the split is accepted at significance level $\varepsilon$ if
\begin{align}\label{eq: threshold U_max}
	U_{\max}^{(n)} = \frac{S - (S_{\leq r^*} + S_{>r^*})}{S/n} > u_{\varepsilon},
\end{align}
where ``$*$'' indicates that we consider the optimal split, and where $u_{\varepsilon}$ is the solution to
\begin{align}\label{eq: threshold tail probability}
	dp_n(u_{\varepsilon}) = \varepsilon,
\end{align}
where $p_n(u)$ is from \eqref{eq: p-value approximation U_j}. An equivalent rephrasing of \eqref{eq: threshold U_max} is
\begin{align}\label{eq: regularisation}
	\RSS_1 - \RSS_2 - u_{\varepsilon} \widehat\sigma^2 > 0,
\end{align}
where $\RSS_1 := S, \RSS_2 := S_{\leq r^*} + S_{>r^*}$, together with $\widehat\sigma^2 := S / n$. A natural question, which is partially answered below, is how $u_{\varepsilon}$ depends on $n$ for a fixed significance level $\varepsilon$. 

\begin{proposition}\label{thm: percentile bound} 
$u_{\varepsilon}$ solving \eqref{eq: threshold tail probability} satisfies $u_{\varepsilon} = o(\ln_2(n))$ as $n\to\infty$. 
\end{proposition}
\noindent The proof of Proposition~\ref{thm: percentile bound} is given in the Appendix.

\medskip

Based on \eqref{eq: regularisation} it is seen that $u_{\varepsilon}$ can be thought of as a regularisation term (or penalty), and from Proposition~\ref{thm: percentile bound} it is seen that this term behaves almost like a constant. We will continue with a short comparison with other techniques that can be used to decide on accepting a split or not.

\subsection{Cost-complexity pruning}\label{sec: cost-complexity pruning}

Cost-complexity pruning was introduced in \cite{breiman1984classification} and is described in terms of the so-called ``cost'' w.r.t.\ a split tolerance $\vartheta$, denoted by $R_\vartheta(T)$, defined as
\begin{align}
	R_\vartheta(T) := R(T) + \vartheta |T|,
\end{align}
where, in our sitting, we have $R(T) = \sum_{i = 1}^n (Y^{(i)}-T(X^{(i)}))^2$ (other loss functions may be considered). The parameter $\vartheta$ is also referred to as the ``cost-complexity'' parameter.
Note that the critical $\vartheta$ value needed in order to accept $T_2(X)$ in favour of $T_1(X)$ is the threshold value $\vartheta$ for which the so-called ``gain'' $R_\vartheta(T_1) - R_\vartheta(T_2)$ is 0, which gives 
\begin{align*}
	R_\vartheta(T_1) - R_\vartheta(T_2) &= R(T_1) - R(T_2) + \vartheta (|T_1| - |T_2|) \\
	&= \RSS_1 - \RSS_2 - \vartheta = 0,
\end{align*}
or equivalently, the split is accepted if $\RSS_1 - \RSS_2 - \vartheta > 0$. The choice of $\vartheta$ used in applications is typically based on out-of-sample performance using, e.g., cross-validation; also recall the discussion in relation to \eqref{eq: CART threshold} above. Using the specific choice $\vartheta := u_{\varepsilon} \widehat\sigma^2$ is equivalent to using the $p$-value based penalty from \eqref{eq: regularisation}. Note that this equivalence only applies to the situation concerning whether one should accept a single split or not, whereas, as mentioned above, the cost-complexity pruning is a procedure that evaluates entire subtrees.

\subsection{Covariance penalty and information criteria}

Another alternative is to assess a candidate split based on its predictive performance using the mean squared error of prediction (MSEP), conditioning on the observed covariate values. When working with linear Gaussian models this corresponds to using Mallows' $C_p$, where $p$ corresponds to the number of regression parameters, see e.g.~\cite{mallows1973comments}, which is an example of an estimate of the prediction error using covariance based penalty, see e.g.\ \cite{efron2004estimation}. The $C_p$ statistic can then be expressed as
\begin{align*}
	C_p := \frac{1}{n}\big(\RSS_p + 2 p \widehat\sigma^2 \big),
\end{align*}
which is the formulation used in \cite[Ch. 7.5, Eq. (7.26)]{hastie2009elements}. Consequently, since a binary single-split $L^2$ regression tree with predetermined split point can be interpreted as fitting a Gaussian model with a single binary covariate, $C_p$ can in this situation be used to evaluate predictive performance. By considering the $C_p$ improvement when going from no split, i.e.~$p = 1$, to one split, $p = 2$, corresponds to $C_1 - C_2 > 0$, which is equivalent to
\begin{align*}
	\RSS_1 - \RSS_2 - 2 \widehat\sigma^2 > 0.
\end{align*}
Thus, using Mallows' $C_p$, targeting the predictive performance of the estimator, will be asymptotically too liberal compared to the $p$-value based stopping rule. This, however, should not be too surprising, since the above application of the $C_p$ statistic does {\em not} take into account that the candidate split point has been chosen by minimising an $L^2$ loss.

For a $p$-parameter Gaussian model the $C_p$ statistic coincides with the Akaike information criterion (AIC), see e.g.~\cite[Ch. 7.5, Eq. (7.29)]{hastie2009elements}.
For a $p$-parameter Gaussian model, the Bayesian information criterion (BIC) considers the quantity  
\begin{align*}
	\BIC_p := \frac{\RSS_p}{\sigma^2} + \ln(n) p, 
\end{align*}
see e.g.~\cite[Ch. 7.7, Eq. (7.36)]{hastie2009elements}, as the basis for model selection. In practice $\sigma^2$ is replaced by a suitable estimator, $\widehat\sigma^2$, see, e.g., the discussion in the paragraph following \cite[Ch. 7.7, Eq. (7.36)]{hastie2009elements}. Hence, it follows that accepting a split based on BIC-improvement in a single split corresponds to $\BIC_1 - \BIC_2 > 0$, which is equivalent to
\begin{align*}
	\RSS_1 - \RSS_2 - \ln(n) \widehat\sigma^2>0.
\end{align*}
Thus, using BIC as a stopping criterion is more conservative than the $p$-value based stopping criterion, despite not taking into account that the split point is given as a result of an optimisation procedure.

\section{Numerical illustrations}\label{sec: numerical illustrations}

All code used in the numerical illustrations is available on GitHub\footnote{See footnote~\ref{fnote: github}.}.

\subsection{The $p$-value approximation for a single split}

In this section we investigate the error from applying the two approximations in  \eqref{eq: p-value approximation U} and \eqref{eq: p-value approximation U_j}. Both together provide the $p$-value approximation used to test for signal. Since we do not have access to the true distribution of $U_{\max}$ under $H_0$, we compute its empirical distribution from 10,000 realisations in order to compare to the approximations.

\begin{figure}[htp!]
\centering
\includegraphics[width=.49\textwidth]{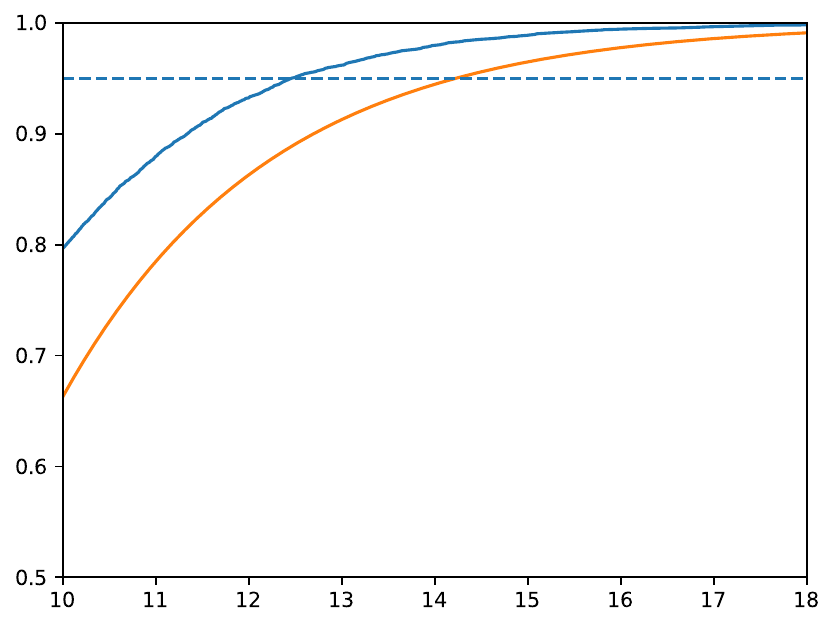}\hfill
\includegraphics[width=.49\textwidth]{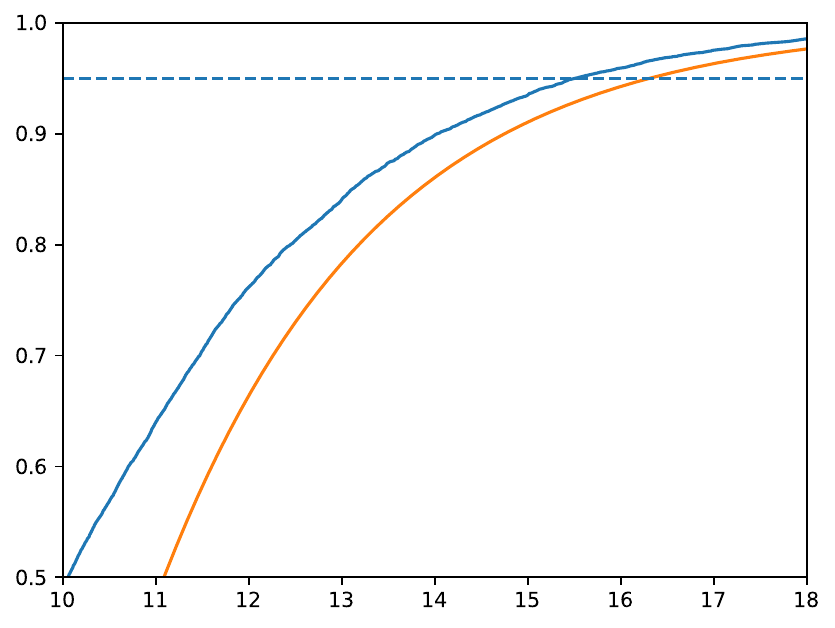}
\includegraphics[width=.49\textwidth]{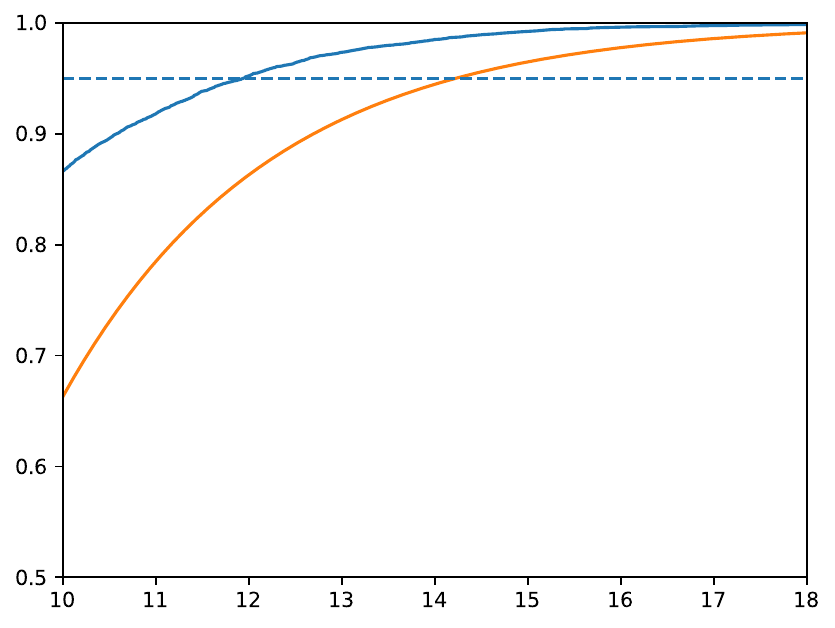}\hfill
\includegraphics[width=.49\textwidth]{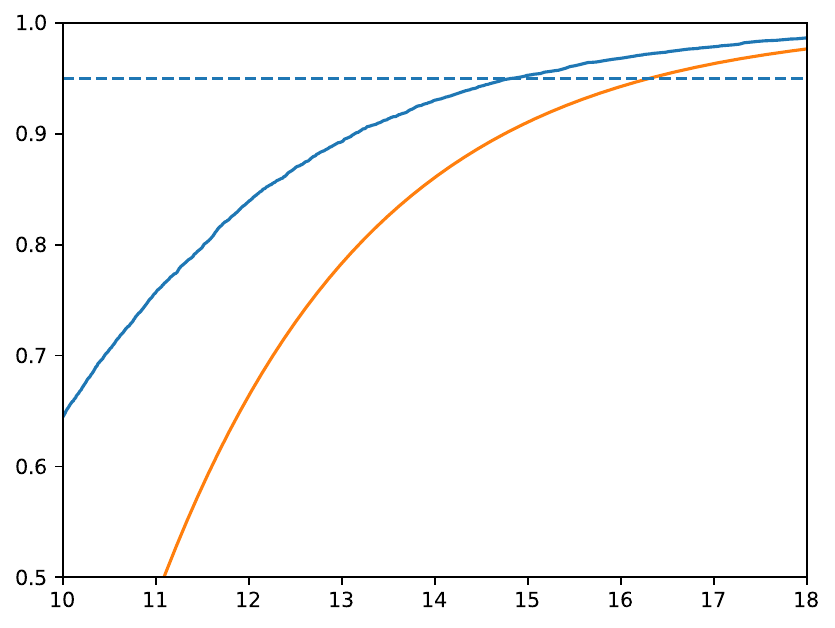}
\caption{Blue curves: empirical cdf of $U_{\max}$ given $H_0$ computed from 10,000 realisations. Orange curves: Approximation $1 - dp_n(u)$. Left column: $n= 50$, right column: $n=$ 1,000. Top row: independent standard normal covariates, bottom row: dependent normal covariates with common pairwise correlation $\rho = 0.8$ and unit variance. The points of intersection with the dashed blue line illustrate empirical and approximate $0.95$-quantile of $U_{\max}$.}
\label{fig:cdf_approx_Yao}
\end{figure}

Figure \ref{fig:cdf_approx_Yao} shows the approximated and true cdfs for varying sample size and covariate dependence. Here, the covariate dimension is set to $d=10$.
Table \ref{fig:cdf_approx_Yao} compares the approximated and true critical quantile values at a $0.95$-level for varying sample size, covariate dimension and covariate dependence. Note that for $d=1$, varying dependence is not an issue so that the entries of the first two tables are identical.
\begin{table}[h!]
\caption{Left table: $0.95$-level quantiles based on the empirical cdf of $U_{\max}$ given $H_0$ computed from 10,000 realisations for independent standard normal covariates. Middle table: The analogous quantiles for dependent normal covariates with a common pairwise correlation of $\rho = 0.8$ and standard variances. Right table: Quantile approximation corresponding to \eqref{eq: p-value approximation U_j}. }
\label{tab:comp_quantiles_bonf}
\centering
\[
\begin{array}{c||c|c}
 & n=50  & n=1,000   \\
\hline\hline
d=1& 8.55 & 10.78    \\
\hline
d=2& 9.79 & 12.10    \\
\hline
d=10& 12.46 & 15.51   \\
\end{array}
\quad
\begin{array}{c|c}
  n=50  & n=1,000   \\
\hline\hline
 8.55 & 10.78    \\
\hline
 9.62 & 12.00    \\
\hline
 11.94 & 14.84   \\
\end{array}
\quad
\begin{array}{c|c}
  n=50  & n=1,000   \\
\hline\hline
 9.12 & 11.09    \\
\hline
 10.67 & 12.68    \\
\hline
 14.23 & 16.31   \\
\end{array}
\]
\end{table}

As was noted in \cite{yao1986asymptotic}[Remark 2.3], the approximation \eqref{eq: p-value approximation U_j} yields satisfactory results even for small sample sizes $20 \leq n \leq 50$. This is confirmed by the first row of Table \ref{tab:comp_quantiles_bonf}. The second row of Figure \ref{fig:cdf_approx_Yao} as well as the middle part of Table \ref{tab:comp_quantiles_bonf} show that a strong positive pairwise correlation of $\rho = 0.8$ between covariates does not substantially affect the upper tail of the distribution of $U_{\max}$ under $H_0$ and that the quantile approximations provide good upper bounds.

We now turn to assuming that the alternative hypothesis $H_A$ according to Definition \ref{def:alt_hyp} holds. In order to illustrate Proposition \ref{thm:pvaluetozero}, we pick $\varepsilon = 0.05$, $\sigma^2 = 1$, $j=1$, $\xi = 0$, $t_0 = 1/2$, $\mu_l = 0$ and  $\mu_r = n^{-1/5}$. Note that the step size is chosen to decrease slowly enough  towards zero in order to fulfil the assumptions of $H_A$ in Definition \ref{def:alt_hyp}.

\begin{figure}[htp!]
\centering
\includegraphics[width=.49\textwidth]{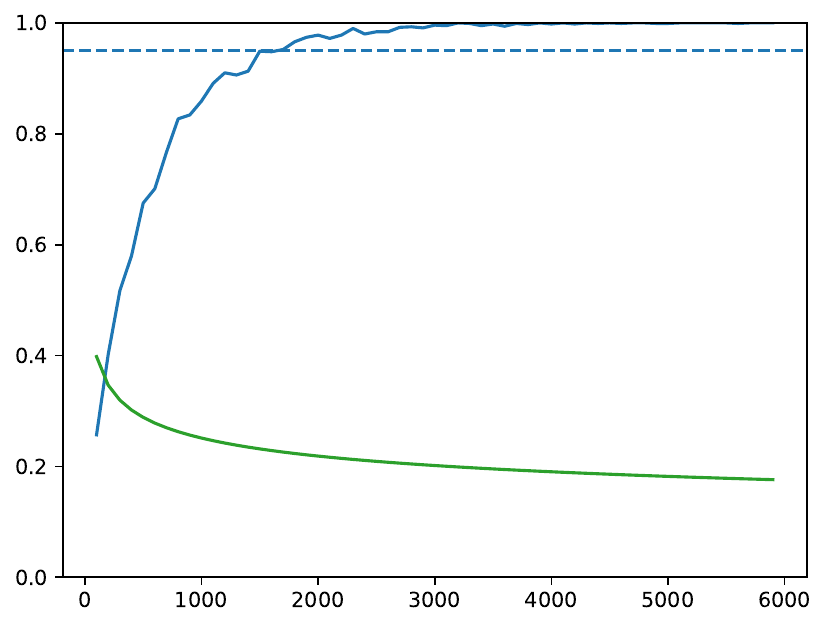}\hfill
\includegraphics[width=.49\textwidth]{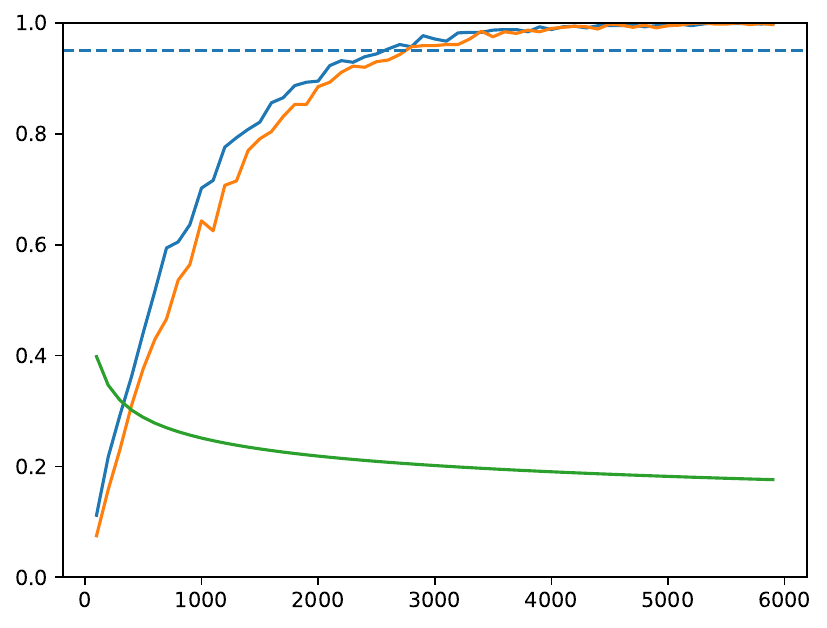}
\caption{Blue curves: Fraction of correct signal detections according to $\{U_{\max}^{(n)} > u_\varepsilon \}$ for an increasing number of data points $n$ and independent standard normal covariates. Orange curve: The analogous fraction based on dependent multivariate normal covariates with common pairwise correlation $\rho = 0.8$ and unit variance. Green curve: The signal strength $|\mu_r-\mu_l| = n^{-1/5}$. The blue dashed line shows the $0.95$-level. The left and right plots correspond to $d=1$ and $d=10$ covariates, respectively.}
\label{fig:typeIId10}
\end{figure}

In Figure \ref{fig:typeIId10}, we plot the fraction of correct signal detections from 1,000 realisations of the event $\{U_{\max}^{(n)} > u_\varepsilon \}$, where $u_\varepsilon$ is given in \eqref{eq: threshold tail probability}. We run the simulations for an increasing number of data points $n$. Figure \ref{fig:typeIId10} confirms the findings of Proposition \ref{thm:pvaluetozero} that the probability of detecting a slowly decreasing signal converges to one as $n$ tends to infinity.

It can be noted that the upper tail of $U_{\max}^{(n)}$ is not affected much by introducing dependence between the covariates, as the orange and blue curves in the right plot of Figure \ref{fig:typeIId10} differ little.

\subsection{Simulated examples from Neufeldt et al.}\label{sec: neufeld}

In this section we fix a simple tree and then generate residuals around its level values in order to illustrate the detection performance of our method. We consider the following example as proposed by \cite[section 5]{neufeld2022tree}. Consider independent standard normal covariates and a regression function given by
\begin{align}\label{eq:mu_neufeldt}
    \mu(x) = b \Big(\mathds{1}_{\{x_1 \leq 0\}} \big(1 +  a\mathds{1}_{\{x_2 > 0\}} + \mathds{1}_{\{x_2 x_3 > 0\}}\big)\Big),
\end{align}
for $x \in \R^{10}$ and parameters $a, b  \in \R$ determining the step size between the level values (signal strength). The step size between siblings at level two is $ab$ while the step size between siblings at level three is $b$. An illustration of the tree corresponding to \eqref{eq:mu_neufeldt} is given in Figure \ref{fig:neufeld_tree}. We generate $500$ iid covariate vectors $X_1, \dots, X_{500}$ of $N(0, I_{10})$ and corresponding response variables $Y_1, \dots, Y_{500}$, where, given $X_i$, $Y_i$ is drawn from $N(\mu(X_i), 1)$.
\begin{figure}[htp!]
\centering
\begin{tikzpicture}[
  level 1/.style={sibling distance=60mm, level distance=15mm},
  level 2/.style={sibling distance=40mm, level distance=15mm},
  level 3/.style={sibling distance=20mm, level distance=15mm},
  every node/.style={rectangle, draw=black, rounded corners, align=center, minimum width=10mm},
  edge from parent/.style={draw}
]
\node {$X_1 \leq 0$ \\ $b$}
  child {
  node {$X_2 \leq 0$ \\ $2b$}
    child {
    node {$X_3 \leq 0$ \\ $1.5b$}
    	child{node{$2b$}}
	child{node{$b$}}
    }
    child {node {$X_3 \leq 0$ \\ $2.5b$}
    	child{node{$2b$}}
	child{node{$3b$}}
    }
  }
  child {
  node {$0$}
  };
\end{tikzpicture}
\caption{Regression tree corresponding to \eqref{eq:mu_neufeldt} with $a=1$ adopted from \cite[section 5]{neufeld2022tree}. Each left leaf answers the inequality with ``true''.}
\label{fig:neufeld_tree}
\end{figure}

Using the python package \texttt{sklearn.tree.DecisionTreeRegressor}, we grow a full CART regression tree of maximal depth $4$ with a minimal number of data points per leaf set to $20$. For each tree in the nested sequence of cost-complexity-pruned subtrees (from the root to the fully grown CART regression tree), we compute the in-sample error (MSE) and out-of-sample error (MSEP), where the latter is done using independently generated test data of the same size $n=500$ which was neither used to fit the CART regression tree, nor to compute $p$-values, but serves only as a data set for pure out-of-sample testing.

\begin{figure}[htp!]
\centering
\includegraphics[width=.49\textwidth]{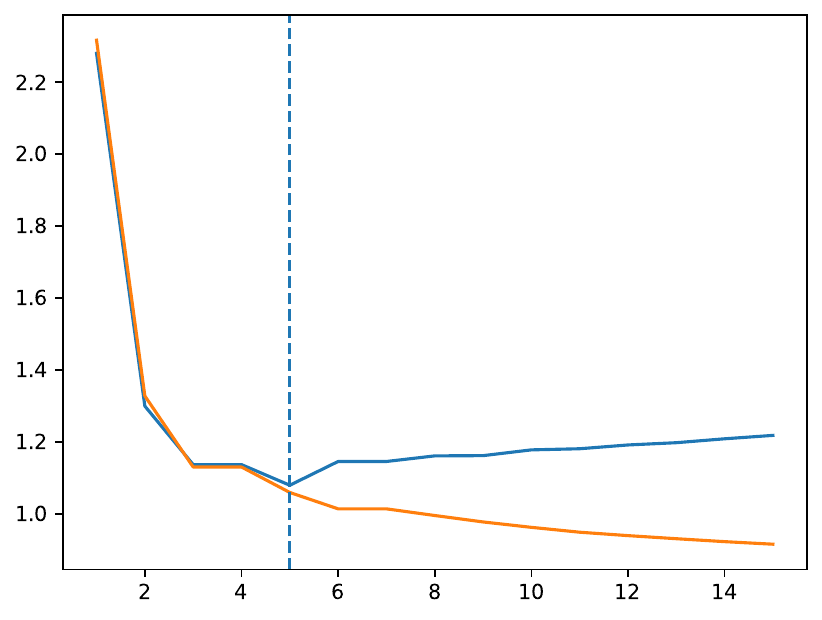}\hfill
\includegraphics[width=.49\textwidth]{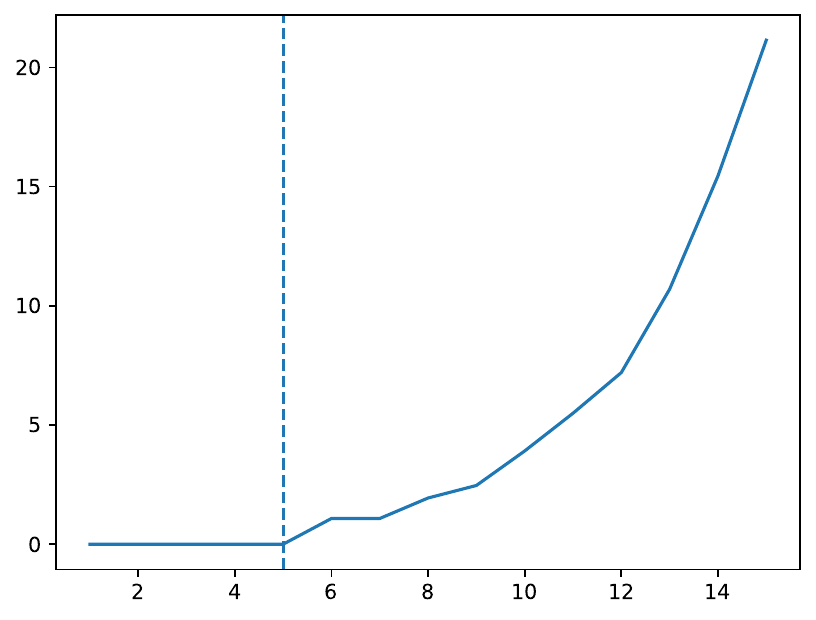}
\caption{Left plot: MSEP (blue) and MSE (orange) for each tree in the nested sequence of cost-complexity-pruned subtrees. Right plot: cumulative $p$-value for each tree in the nested sequence of cost-complexity-pruned subtrees. The $x$-axis depicts the number of leaves of the subtree considered. The dashed blue line marks our method's output tree, i.e. the largest subtree whose cumulative $p$-value lies below $\delta = 0.05$. The signal strength parameters are $a=b=1$.}
\label{fig:MSEMSEP_neufeld}
\end{figure}

In the example of Figure \ref{fig:MSEMSEP_neufeld}, the proposed method detects the correct complexity of $\mu$ which is given by $5$ leaves and which minimises MSEP.  The cumulative $p$-values of all smaller subtrees are very close to zero ($0, 0.0001, 0.0003$), while jumping sharply to $1.08$ after the first ``unnecessary'' split (cf. Figure \ref{fig:output_trees_neufeld}). The results in this example are hence not sensitive to the choice of the tolerance parameter $\delta$. Note that individual $p$-values may exceed one due to the approximation \eqref{eq: p-value bound}. Comparing Figure \ref{fig:neufeld_tree} with the upper tree of Figure \ref{fig:output_trees_neufeld}, we note that also the split points and mean values are accurate.

\begin{figure}[htp!]
\centering
\includegraphics[width=.49\textwidth]{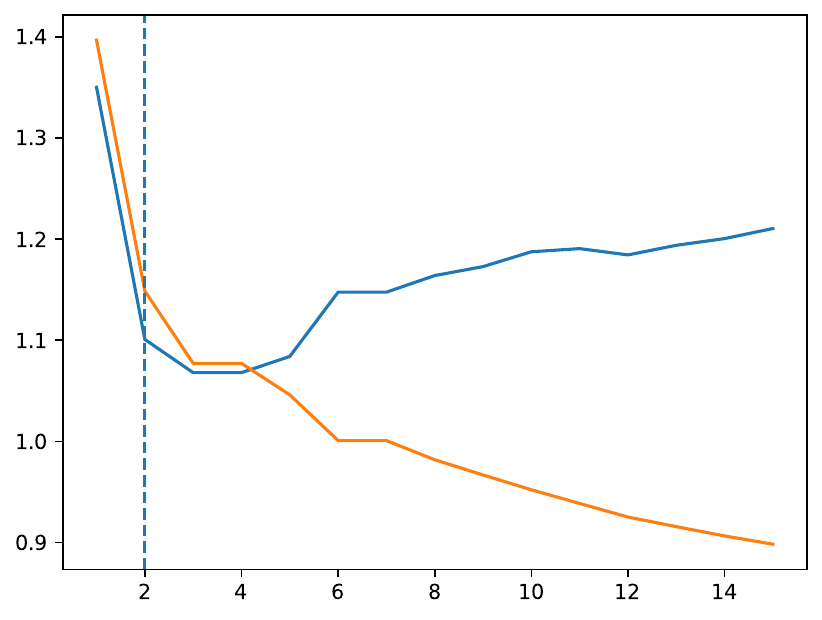}\hfill
\includegraphics[width=.49\textwidth]{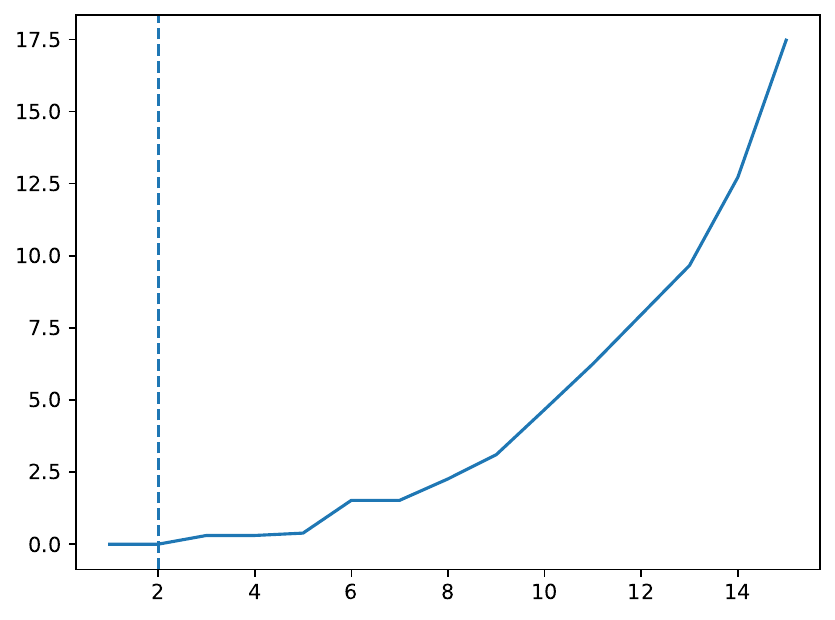}
\caption{Analogue of Figure \ref{fig:MSEMSEP_neufeld} with $b=0.5$ instead of $b=1$.}
\label{fig:MSEMSEP_neufeld_stn05}
\end{figure}

We repeat the simulation for a decreased signal parameter $b=0.5$, while keeping $a=1$, $\sigma^2 = 1$ and $n=500$. As can be observed in  Figure \ref{fig:MSEMSEP_neufeld_stn05} and the bottom tree of Figure \ref{fig:output_trees_neufeld}, the method stops after already one split not capable of detecting the weak signal in the lower part of the tree. However, it regularises well in the sense that MSEPs are close to minimal. Even though the sample size $n=500$ is chosen rather small, the results of Figures \ref{fig:MSEMSEP_neufeld} and \ref{fig:MSEMSEP_neufeld_stn05} do not vary much between runs with different random seeds for the training and validation data generation. These results are primarily a consequence of small sample size, see the discussion in Section~\ref{sec:change_point_detection} on the theoretical guarantees for rejecting a false $H_0$ and for the signal strength corresponding to $H_A$.


\begin{figure}[htp!]
\centering
    \begin{tikzpicture}[ scale = 0.9,
  level 1/.style={sibling distance=80mm, level distance= 30mm},
  level 2/.style={sibling distance=60mm, level distance=30mm},
  level 3/.style={sibling distance=30mm, level distance=30mm},
  every node/.style={rectangle, draw=black, rounded corners, align=center, minimum width=8mm},
  edge from parent/.style={draw}
]
\node {$X_1 \leq 0.00$ \\ $0.99$ \\ $\mathbf{0}$ \\ $\mathbf{0}$}
  child {
  node {$X_2 \leq -0.01$ \\ $2.03$ \\ $\mathbf{8 \cdot 10^{-5}}$ \\ $\mathbf{1 \cdot 10^{-4}}$}
    child {
    node {$X_3 \leq 0.04$ \\ $1.57$ \\ $\mathbf{1 \cdot 10^{-5}}$ \\ $\mathbf{1 \cdot 10^{-4}}$}
    	child { node[fill=red!20] {$X_4 \leq 0.17$ \\ $2.10$ \\ $\mathbf{0.86}$ \\ $\mathbf{1.9377}$} 
        }
        child { node[fill=red!20] { $X_7 \leq 0.3$ \\ $0.75$ \\  $\mathbf{1.58}$ \\ $\mathbf{5.49}$} 
        }
    }
    child {
    node {$X_3 \leq 0.05$ \\ $2.46$ \\ $\mathbf{2\cdot 10^{-4}}$ \\ $\mathbf{3 \cdot 10^{-4}}$}
        child { node[fill=red!20] { $X_4 \leq -0.54$ \\ $2.04$ \\ $\mathbf{0.53}$ \\ $\mathbf{2.46}$} 
        }
        child { node[fill=red!20] {$X_2 \leq 0.75$ \\ $3.15$ \\ $\mathbf{5.69}$ \\ $\mathbf{21.14}$} 
        }
    }
  }
  child {
  node[fill=red!20] {$X_7 \leq 0.22$ \\ $0.04$ \\ $\mathbf{0.94}$ \\ $\mathbf{1.08}$}
  };
\end{tikzpicture} \\
\vspace{1cm}
\centering
\begin{tikzpicture}[ scale = 0.9,
  level 1/.style={sibling distance=80mm, level distance=30mm},
  level 2/.style={sibling distance=60mm, level distance=30mm},
  every node/.style={rectangle, draw=black, rounded corners, align=center, minimum width=8mm},
  edge from parent/.style={draw}
]
\node {$X_1 \leq 0.00$ \\ $0.51$ \\ $\mathbf{0}$ \\ $\mathbf{0}$}
  child {
  node[fill=red!20] {$X_2 \leq -0.01$ \\ $1.03$ \\ $\mathbf{0.28}$ \\ $\mathbf{0.30}$}
     }
  child {
  node[fill=red!20] {$X_7 \leq 0.22$ \\ $0.04$ \\ $\mathbf{0.99}$ \\ $\mathbf{1.52}$}
  };
\end{tikzpicture}
    \caption{Regularised output trees for $b=1$ (top) and $b=0.5$ (bottom). First row of each node: split point selected by CART. Second row: mean value. Third row: node $p$-value. Fourth row: cumulative $p$-value of the smallest subtree the node appears in as a non-leaf. Nodes shaded red violate the condition that the cumulative $p$-value lies below $0.05$.}
    \label{fig:output_trees_neufeld}
\end{figure}

Moreover, from Figure \ref{fig:typeIId10} in Section \ref{sec: neufeld} we can observe that a larger number of data points of around $n=$ 2,500 would ensure (with a $95$ percent probability) the detection of an even lower signal $0.21 < b = 0.5$ in each split of the tree. We conclude that $n=500$ is insufficient in this example with $b=0.5$.

\subsubsection{Illustrating the randomness of tree construction using cross-validation}\label{sec: tree randomness}

In order to illustrate the randomness in trees constructed using cross-validation we simulate data according to a Gaussian linear model. That is, we do not assume that the true model is a tree, which is assumed in the simulation used in \cite{neufeld2022tree}. The data is generated according to the following: the covariates $X \in \mathbb{R}^p$, with $p = 6$, are assumed to be i.i.d.~$N(0, 1)$, and the true regression model is given by
\[
	Y = \beta^{\tr}X + \varepsilon,
\]
where $\varepsilon \sim N(0, 1)$, and where
\[
	\beta :=( 0.517, 0.419 , 1.107, -0.161,-0.913,  -0.984)^\tr.
\]
Based on this model we have simulated $n = 500$ i.i.d.~outcomes of $(Y, X)$, that have been split into a training set of 400 data points and a test set of 100 data points. Using the 400 training data points,  we fit a single cost-complexity pruned tree using 5-fold cross-validation, with at most depth of 8 and min samples per leaf set to 20, together with the corresponding tree fitted using the $p$-value method from the present paper. That is, the cost-complexity pruning procedure produces a sequence of $\alpha$-values, and the $\alpha$-value picked corresponds to the one that minimises the $k$-fold $\operatorname{RMSE}_\alpha$ given by
\begin{align}\label{eq: k-fold RMSE}
	\operatorname{RMSE}_\alpha := \frac{1}{k} \sum_{j = 1}^k \operatorname{RMSE}_{\alpha, j},
\end{align}
where $\operatorname{RMSE}_{\alpha, j}$ is the test RMSE when using the $j$th fold as test set. Further, in order to assess the randomness due to cross-validation, the 5-fold cross-validation procedure is repeated in total 500 times, based on the same initial 400 training data points, by re-sampling 5 new folds in each repetition. This generates 500 potentially different trees, whereas the $p$-value method produces the same tree for a given level of significance level $\delta$.  

\begin{figure}[htp!]
\centering
\includegraphics[width=\linewidth]{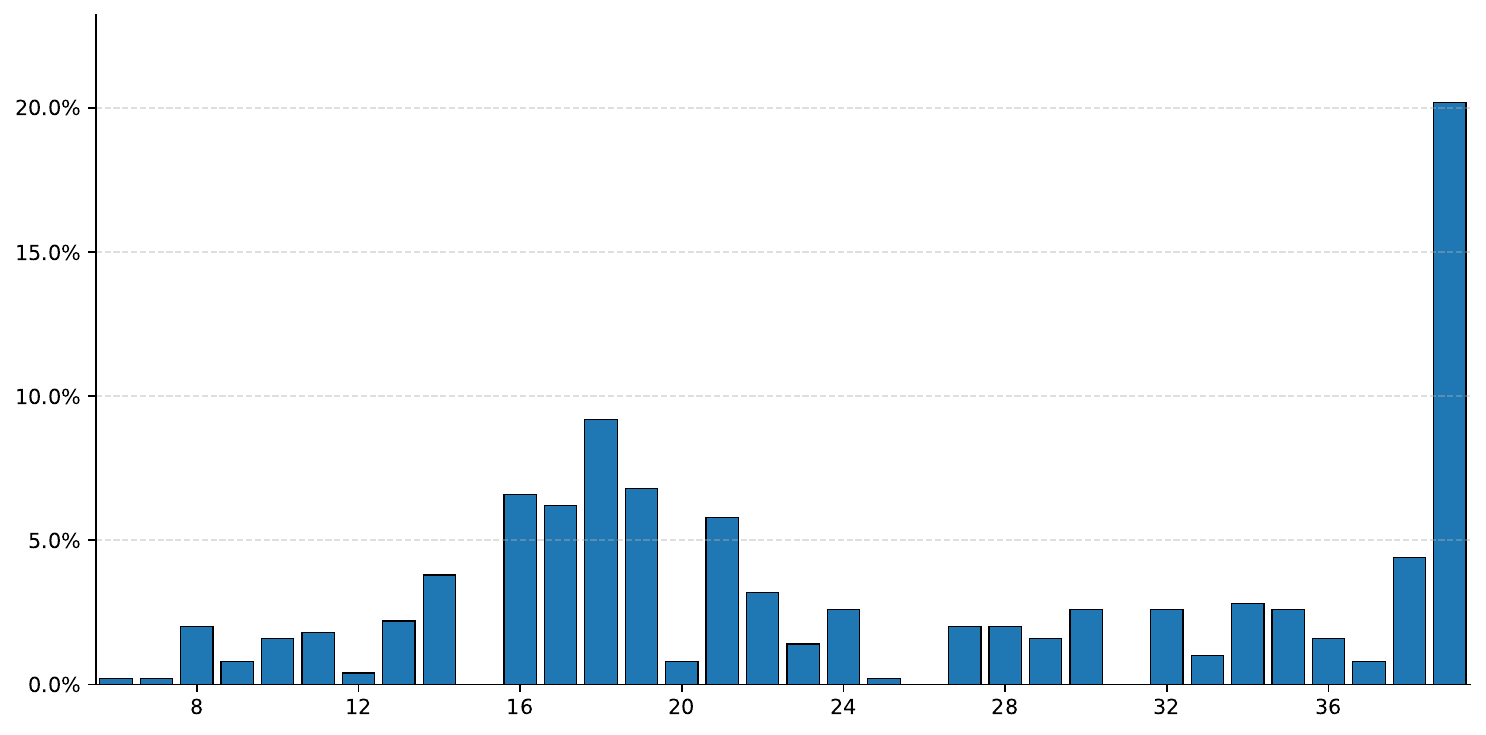}
\caption{Frequencies of number of leaves when fitting an $L^2$-tree using 5-fold cross validation to data simulated as described in Section~\ref{sec: tree randomness}, based on $n = 400$ data points and 500 re-sampled folds.}
\label{fig: leaf frequencies CV}
\end{figure}

In Figure~\ref{fig: leaf frequencies CV} we see how the number of leaves varies between the different cross-validation runs. The tree with the lowest test RMSE value corresponds to a cross-validation based tree with 19 leaves and a test RMSE of $1.5635$. This can be compared to the $p$-value based trees seen in Table~\ref{tab: cv pval}, whose test RMSE values are comparable to the optimal cross-validation pruned tree. In particular, from Figure~\ref{fig: leaf frequencies CV} one can note that the optimal cross-validation pruned tree is not the most likely tree. Moreover, compared to the $p$-value based trees from Table~\ref{tab: cv pval}, the cross-validation pruned trees will likely be much more complex, noting that approximately 20\% of these trees will have 39 leaves compared to about 10 leaves for the $p$-value based trees.

As a final comment, from \eqref{eq: k-fold RMSE} it is clear that the cross-validation procedure may become very time consuming when using larger data sets.

\begin{table}[h!]
\caption{Performance in terms of test RMSE of trees obtained using the $p$-value method when using  different $\delta$-values.}
\label{tab: cv pval}
\centering
\begin{tabular}{ccc}
\hline
$\delta$ & Leaves & RMSE \\
\hline
0.01 & 8 & 1.6697 \\
0.05 & 11 & 1.6329 \\
0.10 & 12 & 1.5973 \\
\hline
\end{tabular}
\end{table}

\subsection{An application of creating an auto-calibrated predictor using $L^2$ trees}\label{sec: auto-calibration L2}

In applications it is rarely the case that one wants to build a single large tree directly from data, due to, e.g., stability concerns and lack of predictive performance, see e.g.~\cite[Ch.~9.2]{hastie2009elements}. Trees are, however, typically used as base learners in ensemble methods such as (gradient) boosting and random forests. In the current paper we do not explore this further, but note that the suggested $p$-value approach can be used as an automatic deterministic stopping criterion in, e.g., boosting, which is similar to the ABT-machine from \cite{huyghe2024boosting}.

Another application is to approximate a complex ``black-box'' model by a single regression tree. One aspect of this is so-called model distillation, see e.g.~\cite{zhou2024approximation}, which uses a related $p$-value based tree approach. In this paper we instead consider how to use a single tree to obtain an auto-calibrated locally unbiased version of a given more complex black-box predictor, see e.g.~\cite{kruger2021generic}.

The idea behind auto-calibration is as follows: Let $\widehat\mu(X)$ denote a given predictor, and let
\begin{align}\label{eq: auto-calibration}
	\widehat\mu^*(X) := \E[Y \mid \widehat\mu(X)].
\end{align}
A predictor $\widehat\mu(X)$ is said to be auto-calibrated if $\widehat\mu^*(X) = \widehat\mu(X)$ a.s., see e.g.~\cite{kruger2021generic}. It is clear that not all predictors $\widehat\mu(X)$ are auto-calibrated, but based on \eqref{eq: auto-calibration} and the tower property of iterated conditional expectations, it follows that
\begin{align*}
	\E[Y \mid \widehat\mu^*(X)] = \E[Y \mid \E[Y \mid \widehat\mu(X)]] 
		= \E[Y \mid \widehat\mu(X)]
		= \widehat\mu^*(X).
\end{align*}
Consequently, by defining the predictor $\widehat\mu^*(X)$ according to \eqref{eq: auto-calibration} it is, in theory, possible to construct an auto-calibrated predictor $\widehat\mu^*(X)$, given any predictor $\widehat\mu(X)$. Construction of a piece-wise constant auto-calibrated predictors is discussed in \cite{lindholm2023local, wuthrich2024isotonic}, corresponding to letting the original predictor $\widehat\mu(X)$ provide guidance on how to partition the covariate space. The question then is how to create this partition. In \cite{lindholm2023local} a simple $L^2$-tree procedure is used, but as pointed out in \cite{wuthrich2024isotonic}, this procedure has an inherent element of randomness in itself due to the use of cross-validation. Thus, by using the $p$-value based method discussed in the present paper we obtain an automatic deterministic procedure for constructing a partition and the corresponding auto-calibrated piece-wise constant predictor. For more on this, see e.g.~\cite{lindholm2023local}.

To demonstrate this procedure we consider the California Housing data\footnote{\label{fnote: cal housing}Available at \texttt{https://www.kaggle.com/datasets/dhirajnirne/california-housing-data}}, which has $n=$~20,640 data points, and $d=8$ covariates (removing the categorical covariate ``ocean proximity''\footnote{See footnote \ref{fnote: cal housing}.}), where we use median house value as our response variable $Y$.

As our initial black-box predictor we use a gradient boosting machine (GBM), see \cite{friedman2001greedy}, which minimises the $L^2$ loss  based on 5-fold cross-validation, where 20\% of data is removed for test performance evaluation. The max depth of the weak learners in the GBM is set to $3$, i.e.~a tree with at most 8 leaves can be added in a single iteration, the minimum samples per leaf is set to $20$, and we set the learning rate for the boosting procedures to $0.1$. The resulting predictor, denoted $\widehat\mu^{\text{GBM}}(x)$, is trained for approximately 2,500 iterations before stopping. One could consider tuning the shrinkage parameter in order to adjust the number of boosting steps, but this has not been investigated further in the present paper.

\begin{table}[h!]
\centering
\caption{Results for California Housing. Complexity corresponds to the number of trees for the GBM and for the number of leaves for the single tree models. For the trees, ``raw'' corresponds to a tree built from the original data, whereas ``BBG'' corresponds to a ``black-box guided'' (here GBM) tree.}\label{table: MSEP real data}
\begin{tabular}{llrrr}
\hline
 \textbf{Model} && \textbf{RMSE train} & \textbf{RMSE test} & \textbf{Complexity} \\
\hline
Intercept && 1.156 & 1.145 & ---~~ \\
GBM && 0.324 & 0.453 & 2,282 \\
Tree, raw &$\delta = 0.05$ & 0.613 & 0.652 & 83 \\
Tree BBG, &$\delta = 0.05$ & 0.318 & 0.454 & 27 \\
\hline
\end{tabular}
\end{table}

From Table~\ref{table: MSEP real data} we see that the test data performance of the GBM provides a considerable improvement compared to an intercept only model, which clearly implies that the GBM has found signal in the data. Moreover, in Table~\ref{table: MSEP real data} we have also added the results for fitting a single large $L^2$-tree using the $p$-value method, which provides a clear improvement compared to the  intercept only model, but does not come close to the GBM performance.

Continuing, given the fitted GBM, $\widehat\mu^{\text{GBM}}(x)$, we introduce the new one-dimensional real-valued covariate $x_i^*$ defined according to
\[
	x_i^* := x_i^*(x_i) := \widehat\mu^{\text{GBM}}(x_i),
\]
and let $(y_i, x_i^*)$, $i = 1, \ldots, n$, denote the data to be used to create an auto-calibrated tree predictor. That is, $x^* \in \mathbb{R}$ is a non-trivial function of the original $x \in \mathbb{R}^d$ via the GBM-predictor, and hence a tree predictor fitted to $(y_i, x_i^*)$, $i = 1, \ldots,n$, is a non-trivial function in terms of the original $x$. The resulting auto-calibrated ``black-box guided'' tree will be referred to as ``Tree BBG''.

\begin{figure}[htp!]
\centering
\includegraphics[width=\linewidth]{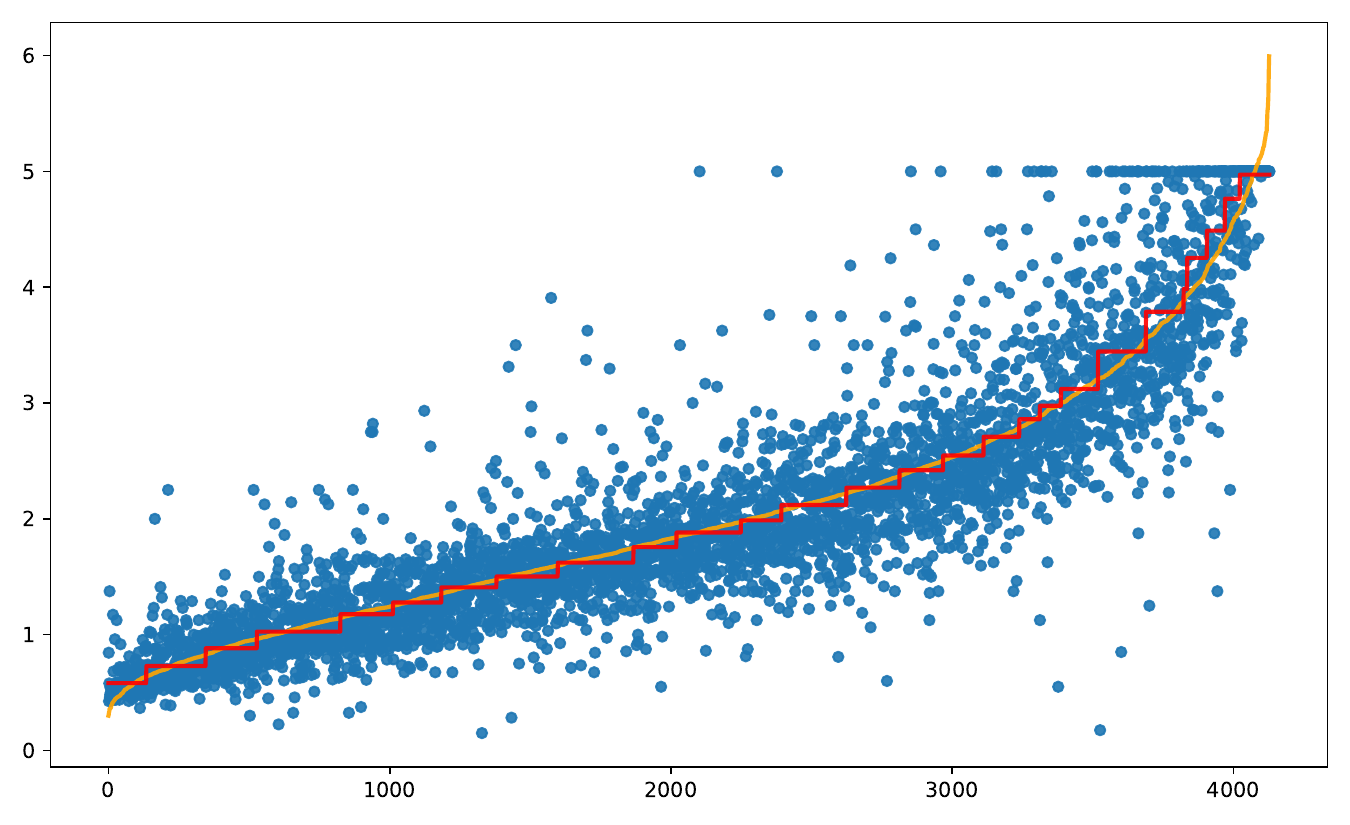}
\caption{Response values for the California Housing data ($y$-axis) as a function of ranks based on the ordering of the $\widehat\mu^{\text{GBM}}(x)$ on the test set ($x$-axis). Orange line corresponds to the original GBM-predictor and the red line corresponds to the auto-calibrated $L^2$-tree fitted to the GBM predictor, i.e.~Tree BBG, as described in Section~\ref{sec: auto-calibration L2}.}
\label{fig: tree and GBM}
\end{figure}

The test RMSE for the resulting Tree BBG, i.e.~the $L^2$-tree fitted to $(Y, X^*)$-data using the $p$-value approach with $\delta = 0.05$, is given in Table~\ref{table: MSEP real data}, where it is seen that by using a single $L^2$-tree with 27 leaves we produce a test error that is very close to the initial GBM that uses approximately 2,500 trees of depth at most 3. Further, in Figure~\ref{fig: tree and GBM} we plot the test $y_i$s against the ranks induced by the ordering $x_{(1)}^* < x_{(2)}^* < \ldots < x_{(n)}^*$, together with the GBM predictions and the Tree BBG predictions. From Figure~\ref{fig: tree and GBM} it is clear that the Tree BBG performs well and seems to avoid spurious extrapolation seen in the original GBM. Furthermore, the Tree BBG predictor is a deterministic auto-calibrated predictor, and this predictor avoids the monotonicity assumption needed for producing an auto-calibrated piece-wise constant deterministic predictor using isotonic regression, see e.g.~\cite{wuthrich2024isotonic}.

\bibliographystyle{apalike}
\bibliography{refs}

\appendix

\section{Proofs}\label{sec: proofs}

\subsection{Proof of Proposition \ref{thm:pvaluetozero}}

Before starting the proof of Proposition \ref{thm:pvaluetozero} we note the following:
\begin{remark}
The distribution of the observed test statistic $U^{(n)}_{\max}$ does not depend on $\sigma$ under the alternative hypothesis.   
Under the alternative hypothesis, for any $r\in\{1,\dots,n\}$ and $b\in\{1,\dots,r\}$ such that $X^{(i)}_j<\xi$ for $i\leq b$ and $X^{(i)}_j\geq \xi$ for $i> b$, we may write 
\begin{align*}
Y^{(i)}=Z^{(i)}+
\left\{\begin{array}{ll}
\mu_l, & i=1,\dots,b,\\
\mu_r, & i=b+1,\dots,n.
\end{array}\right.
\end{align*}
where $Z^{(1)},\dots,Z^{(n)}$ are independent and $N(0,\sigma^2)$ distributed. Then 
$\overline{Y}_{\leq r}=\overline{Z}_{\leq r}+(b\mu_l+(r-b)\mu_r)/r$
and 
\begin{align*}
Y^{(i)}-\overline{Y}_{\leq r}
=Z^{(i)}-\overline{Z}_{\leq r}+
\left\{
\begin{array}{ll}
(\mu_l-\mu_r)(r-b)/r, & i=1,\dots,b,\\
(\mu_r-\mu_l)b/r, & i=b+1,\dots,r.
\end{array}\right.
\end{align*}
Hence, $Y^{(i)}-\overline{Y}_{\leq r}$ equals $\sigma$ times a random variable whose distribution does not depend on $\sigma$. This also holds for $Y^{(i)}-\overline{Y}_{> r}$. We conclude that the distribution of $U^{(n)}_j$ does not depend on $\sigma$ under the alternative hypothesis. 
\end{remark}

\begin{remark}
By construction 
\begin{align}
U^{(n)}_j\frac{S/n}{\sigma^2}=\max_{1\leq r\leq n-1}\frac{1}{\sigma^2}\big(S-S_{\leq r}-S_{>r}\big). \label{eq:Tn2} 
\end{align}
Under the alternative hypothesis, by \cite{yao1986asymptotic} p.~347,   
\begin{align*}
\max_{1\leq r\leq n-1}\frac{1}{\sigma^2}\big(S-S_{\leq r}-S_{>r}\big) 
\eqdis \max_{1\leq nt\leq n-1}\frac{(W_0(t)-f_n(t))^2}{t(1-t)},
\end{align*}
where $W_0$ is a standard Brownian bridge and 
\begin{align*}
f_n(t)=\left\{\begin{array}{ll}
n^{1/2}\theta_n t(1-[nt_0]/n), & \text{if } nt\leq [nt_0], \\
n^{1/2}\theta_n (1-t)[nt_0]/n, & \text{if } nt> [nt_0].
\end{array}\right. 
\end{align*}
\end{remark}  

\begin{proof}[Proof of Proposition \ref{thm:pvaluetozero}]
Since $p_{n}$ is a decreasing function we know that $P_{\max}^{(n)}\leq dp_{n}(U^{(n)}_j)$ for every $j$, in particular for $j$ for which there is signal with amplitude $\sigma\theta_n$ according to the model $\calA^{(n)}$. Hence, 
\begin{align*}
\P_{\calA^{(n)}}(P_{\max}^{(n)}>\varepsilon)
&\leq \P_{\calA^{(n)}}(p_{n}(U^{(n)}_j)>\varepsilon/d)\\
&=1-\P_{\calA^{(n)}}(U^{(n)}_j>p_{n}^{-1}(\varepsilon/d))
\end{align*}
Let $T_n^2$ denote the quantity $U^{(n)}_j(S/n)/\sigma^2$ in \eqref{eq:Tn2}. Then 
\begin{align*}
\P_{\calA^{(n)}}(U^{(n)}_j>p_{n}^{-1}(\varepsilon/d))
=\P_{\calA^{(n)}}\bigg(T_n^2\bigg(\frac{c_n^2}{p_{n}^{-1}(\varepsilon/d)}\frac{\sigma^2}{S/n}\bigg)>c_n^2\bigg) 
\end{align*}
for any positive sequence $(c_n^2)$. We consider the choice of sequence 
\begin{align}\label{eq:cn2}
c_n^2=\bigg(\frac{2^{-1}\ln_3(n)-\ln(2^{-1}\pi^{1/2}\ln((1-\alpha)^{-1}))}{(2\ln_2(n))^{1/2}}+(2\ln_2(n))^{1/2}\bigg)^2 
\end{align}
in order to relate the tail probability $\P_{\calA^{(n)}}(U^{(n)}_j>p_{n}^{-1}(\varepsilon/d))$ to the tail probability $\P_{\calA^{(n)}}(T_n^2>c_n^2)$ studied by \cite{yao1986asymptotic}. 
By Lemma \ref{lem:first_inequality}, 
\begin{align*}
\liminf_{n\to\infty}\P_{\calA^{(n)}}(U^{(n)}_j>p_{n}^{-1}(\varepsilon/d))\geq \liminf_{n\to\infty}\P_{\calA^{(n)}}(T_n^2>c_n^2).
\end{align*}
For any $\eta\in\R$, by Lemma \ref{lem:second_inequality}, 
\begin{align*}
\liminf_{n\to\infty}\P_{\calA^{(n)}}(T_n^2>c_n^2)\geq \alpha+\Phi(\eta)(1-\alpha). 
\end{align*}
Hence, for any $\eta\in\R$, 
\begin{align*}
\limsup_{n\to\infty}\P_{\calA^{(n)}}(P_{\max}^{(n)}>\varepsilon)&\leq \limsup_{n\to\infty}\big(1-\P_{\calA^{(n)}}(T_n^2>c_n^2)\big)\\
&\leq 1-\alpha-\Phi(\eta)(1-\alpha). 
\end{align*}
Since we may choose $\eta$ arbitrarily large, the proof is complete. 
\end{proof}

\begin{lemma}\label{lem:first_inequality}
$\liminf_{n\to\infty}\P_{\calA^{(n)}}(U^{(n)}_j>p_{n}^{-1}(\varepsilon/d))\geq \liminf_{n\to\infty}\P_{\calA^{(n)}}(T_n^2>c_n^2)$
\end{lemma}
\begin{proof}
Let 
\begin{align*}
F_n:=\frac{c_n^2}{p_{n}^{-1}(\varepsilon/d)}\frac{\sigma^2}{S/n}
\end{align*}
and note that 
\begin{align*}
\P_{\calA^{(n)}}(U^{(n)}_j>p_{n}^{-1}(\varepsilon/d))&=\P_{\calA^{(n)}}(T_n^2F_n>c_n^2)\\
&\geq \P_{\calA^{(n)}}(T_n^2F_n>c_n^2 \mid F_n <1)\P_{\calA^{(n)}}(F_n<1)\\
&\quad+\P_{\calA^{(n)}}(T_n^2>c_n^2).
\end{align*}
We will show that $\lim_{n\to\infty}\P_{\calA^{(n)}}(F_n<1)=0$ from which the conclusion follows.  
By Lemma \ref{lem:sequences}, 
\begin{align}\label{eq:pninvsmallerthancn2}
\lim_{n\to\infty}p_{n}^{-1}(\varepsilon/d)/c_n^2=0.
\end{align}
Under $\calA^{(n)}$ there exist independent $Z^{(i)}\sim N(0,\sigma^2)$ and $r\in\{1,\dots,n\}$ such that $Y^{(i)}=Z^{(i)}$ for $i\leq r$, and $Y^{(i)}=Z^{(i)}+\sigma\theta_n$ for $i>r$. Therefore, 
\begin{align*} 
S&=\sum_{i=1}^{r}(Y^{(i)}-\overline{Y}_{\leq n})^2+\sum_{i=r+1}^{n}(Y^{(i)}-\overline{Y}_{\leq n})^2\\
&=\sum_{i=1}^{n}(Z^{(i)}-\overline{Z}_{\leq n})^2+\sigma^2\theta_n^2\bigg(r\bigg(\frac{n-r}{n}\bigg)^2+(n-r)\bigg(\frac{r}{n}\bigg)^2\bigg)\\
&\quad+2\sum_{i=1}^{r}(Z^{(i)}-\overline{Z}_{\leq n})\sigma\theta_n\frac{n-r}{n}+2\sum_{i=r+1}^{n}(Z^{(i)}-\overline{Z}_{\leq n})\sigma\theta_n\frac{r}{n}
\end{align*}
Therefore, by H\"older's inequality applied to the sum of the last two terms above, 
\begin{align*}
\frac{S}{n}&\leq \frac{1}{n}\sum_{i=1}^{n}(Z^{(i)}-\overline{Z}_{\leq n})^2+\sigma^2\theta_n^2+2\bigg(\frac{1}{n}\sum_{i=1}^{n}(Z^{(i)}-\overline{Z}_{\leq n})^2\bigg)^{1/2}\sigma\theta_n\\
&=\bigg(\bigg(\frac{1}{n}\sum_{i=1}^{n}(Z^{(i)}-\overline{Z}_{\leq n})^2\bigg)^{1/2}+\sigma\theta_n\bigg)^2. 
\end{align*}
Since the first term inside the square converges in probability to $\sigma$ and since the second term is bounded we conclude that $\lim_{n\to\infty}\P_{\calA^{(n)}}(F_n<1)=0$. 
The proof is complete. 
\end{proof}

\begin{lemma}\label{lem:second_inequality}
For every $\eta\in\R$, $\liminf_{n\to\infty}\P_{\calA^{(n)}}(T_n^2>c_n^2)\geq \alpha+\Phi(\eta)(1-\alpha)$. 
\end{lemma}
\begin{proof}
Fix $\eta\in\R$. From the expression for the tail probability on page 350 in \cite{yao1986asymptotic} we see that for each $n$, 
\begin{align*}
\P_{\calA^{(n)}}(T_n^2>c_n^2)\geq \P(B_{n,1} \cap B_{n,2} \cup A_n(\theta_n)).
\end{align*}
The events $B_{n,1},B_{n,2}$ are independent of $\theta_n$ and given by  
\begin{align*}
B_{n,1}=\bigg\{\max_{t\in D_{n,1}}\frac{|W(t)|}{t^{1/2}}>c_n\bigg\}, \quad
B_{n,2}=\bigg\{\max_{t\in D_{n,2}}\frac{|W(t)-W(1)|}{(1-t)^{1/2}}>c_n\bigg\},
\end{align*}
where $W$ is standard Brownian motion and $D_{n,1},D_{n,2}$ are index sets. 
The event $A_n(\theta_n)$ is increasing in $\theta_n$ and given by the expression on p.~350 in \cite{yao1986asymptotic} (there with $\theta$ instead of $\theta_n$). Writing $\theta_n=\theta(n,\eta_n)$ for $\theta_n$ in \eqref{eq:stepsize}, note that $\theta(n,\eta_n)\geq \theta(n,\eta)$ for $n$ sufficiently large since $\eta_n\to\infty$ as $n\to\infty$. Hence, for $n$ sufficiently large, 
\begin{align*}
\P_{\calA^{(n)}}(T_n^2>c_n^2)\geq \P(B_{n,1} \cap B_{n,2} \cup A_n(\theta(n,\eta)))
\end{align*}
and the right-hand side converges to $\alpha+\Phi(\eta)(1-\alpha)$ as concluded on p.~350 in \cite{yao1986asymptotic}. 
The proof is complete. 
\end{proof}

\begin{lemma}\label{lem:sequences}
$\lim_{n\to\infty}p_{n}^{-1}(\varepsilon/d)/c_n^2=0$ 
\end{lemma}
\begin{proof}
We have, from the definition of $p_{n}$,  
\begin{align}\label{eq:pninv}
p_{n}^{-1}(\varepsilon/d)=\bigg(\frac{\ln_3(n)+\ln(2)}{(2\ln_2(n))^{1/2}}+\Phi^{-1}\Big((1-\varepsilon/d)^{1/(2\ln(n/2))}\Big)\bigg)^2,
\end{align}
where the first term vanishes asymptotically and the second term tends to $\infty$ as $n\to\infty$.  
Similarly, in \eqref{eq:cn2} the first term vanishes asymptotically and the second term tends to $\infty$ as $n\to\infty$. Hence, it is sufficient to compare the two terms that are not vanishing asymptotically and show that 
\begin{align*}
\lim_{n\to\infty}\frac{\Phi^{-1}(x_n)}{\Phi^{-1}(y_n)}=0, \quad x_n:=(1-\varepsilon/d)^{1/(2\ln(n/2))}, \quad y_n:=\Phi((2\ln_2(n))^{1/2}). 
\end{align*}
By l'Hospital's rule, the convergence follows if we verify that 
\begin{align*}
\lim_{n\to\infty}\frac{\phi(\Phi^{-1}(y_n))}{\phi(\Phi^{-1}(x_n))}=0.
\end{align*}
Note that $\phi(\Phi^{-1}(y_n))=(\sqrt{2\pi}\ln(n))^{-1}\to 0$ as $n\to\infty$. 
The Mill's ratio bound $(1-\Phi(z))/\phi(z)<1/z$ for $z>0$ yields, with $z=\Phi^{-1}(x_n)$, 
\begin{align*}
\phi(\Phi^{-1}(x_n))>\Phi^{-1}(x_n)(1-x_n), \quad x_n>1/2. 
\end{align*}
Hence, 
\begin{align*}
\frac{\phi(\Phi^{-1}(y_n))}{\phi(\Phi^{-1}(x_n))}\leq \frac{1}{\sqrt{2\pi}\ln(n)\Phi^{-1}(x_n)(1-x_n)}. 
\end{align*}
We claim that $\ln(n)(1-x_n)$ converges to a positive limit as $n\to\infty$. Since $\Phi^{-1}(x_n)\to\infty$ as $n\to\infty$ verifying this claim will prove the statement of the lemma. 
Note that 
\begin{align*}
\Big(1-\varepsilon/d\Big)^{1/(2\ln(n/2))}
=\exp\bigg(\frac{\ln\big(1-\varepsilon/d\big)}{2\ln(n/2)}\bigg)
\end{align*}
and hence
\begin{align*}
1+\frac{\ln\big(1-\varepsilon/d\big)}{2\ln(n/2)}&<\Big(1-\varepsilon/d\Big)^{1/(2\ln(n/2))}\\
&<1+\frac{\ln\big(1-\varepsilon/d\big)}{2\ln(n/2)}+\frac{1}{2}\bigg(\frac{\ln\big(1-\varepsilon/d\big)}{2\ln(n/2)}\bigg)^2.
\end{align*}
Hence, with $\gamma:=-\ln(1-\varepsilon/d)$, 
\begin{align*}
\frac{\gamma}{2}\frac{\ln(n)}{\ln(n/2)}>\ln(n)(1-x_n)>\frac{\gamma}{2}\frac{\ln(n)}{\ln(n/2)}-\frac{\gamma^2}{8}\frac{\ln(n)}{\ln(n/2)^2}
\end{align*}
which shows that $\lim_{n\to\infty}\ln(n)(1-x_n)=\gamma/2$. 
The proof is complete. 
\end{proof}

\subsection{Proof of Proposition~\ref{thm: percentile bound}}

\begin{proof}
Note that \eqref{eq: threshold tail probability} is equivalent to $u_{\varepsilon}=p_n^{-1}(\varepsilon/d)$. 
Lemma \ref{lem:sequences} says that $\lim_{n\to\infty}p_{n}^{-1}(\varepsilon/d)/c_n^2=0$. 
Note that $c_n^2$ given by \eqref{eq:cn2} takes the form
\begin{align*}
c_n^2=(d_n+(2\ln_2(n))^{1/2})^2, 
\end{align*} 
where $\lim_{n\to\infty}d_n=0$. The inequality $(a+b)^2\leq 2(a^2+b^2)$ gives 
\begin{align*}
c_n^2=(d_n+(2\ln_2(n))^{1/2})^2\leq 2(d_n^2+2\ln_2(n)). 
\end{align*} 
Hence, $\lim_{n\to\infty}u_{\varepsilon}/\ln_2(n)=0$ which completes the proof. 
\end{proof}

\end{document}